\documentclass[prb,preprint,superscriptaddress]{revtex4-1}
\usepackage[dvips]{epsfig}
\usepackage{bm}
\usepackage{graphicx,amssymb}
\begin{document}

\title{General expressions for the dispersion relation of acoustic radial (breathing) modes in cylinders
and cylindrical shells of general anisotropic crystals}

\bigskip

\author{D. Mart\'{\i}nez-Guti\'errez}
\author{V. R. Velasco}
\affiliation{Instituto de Ciencia de Materiales de Madrid, (ICMM,
CSIC), Sor Juana In\'es de la Cruz 3, 28049 Madrid, Spain}

\begin{abstract}
We present here a study of the radial modes for infinitely long
cylinders and cylindrical shells of general anisotropic crystals. 
The elastic coefficients matrix includes twenty-one independent constants. 
We obtain expressions in closed form for the dispersion relation,
valid for any anisotropic material. In the case of the lowest
breathing mode of a thin cylindrical shell we obtain a simple
analytical formula. This can be used to obtain a first estimate of
the breathing mode frequency in nanotubes for any material.

\end{abstract}

\maketitle

\section{Introduction}

Cylindrical systems are frequently used as structural
components in many engineering areas: aerospace, civil engineering, etc., 
and their vibration characteristics are obviously
important for practical design. 

The study of acoustic wave propagation in infinitely long
homogeneous cylinders and cylindrical layers is a well established
field from the initial studies of Pochhammer \cite{1} and Chree
\cite{2,3}. They developed exact solutions for torsional and
longitudinal vibrations of infinitely long homogeneous isotropic
solid cylinders with stress free faces. Gazis provided solutions
for the acoustic wave propagation problem in homogeneous isotropic
cylinders of infinite length \cite{4}. Studies for orthotropic
hollow and thick cylinders of infinite length, including nine independent 
elastic constants, were done by Mirsky \cite{5}. Detailed discussions on 
these problems and more references can be found in Refs.6-8.

No general expressions for the dispersion relation of the acoustic 
waves in cylinders of general anisotropic crystals
are available, to our knowledge, although it has been possible to obtain
solutions in closed form for some particular kinds of modes and crystal systems. 
The radial (breathing) modes of cylinders have a much simpler form
than a general vibrational mode, thus making them good candidates to 
try to obtain its dispersion relation for general anisotropic crystals.

On the other hand the study of carbon nanotubes (CNT) and
nanowires of many other materials, belonging to different 
crystal systems, has increased the use of
isotropic continuum models for cylinders and cylindrical shells
\cite{9,10,11,12,13}. In the case of transverse elastic isotropy
(hexagonal crystals) formal expressions were given in Ref.14.
In the case of general anisotropic crystals, a method developed in
resonant ultrasound spectroscopy to get the free vibrational modes
of inhomogeneous systems \cite{15,16,17,18}, has been employed in
several studies of acoustic modes in nanowires \cite{19,20,21,22}.

Quite recently the acoustic modes in nanorods of different materials
\cite{23,24,25} have been observed directly by optical methods. 
Other approaches in continuum media \cite{26,27,28} are also been used. This 
shows the interest in the study of the vibrational modes of nanotubes and 
nanowires.

Thus it is worthwile to try to get expressions in closed form for the 
radial modes frequencies in cylinders and cylindrical shells of general 
anisotropic crystals.

We shall consider here an infinitely long cylinder, or cylindrical shell, of
a general anisotropic material and we shall obtain the dispersion relation of the
radial modes for these systems.

In Section II we present the formal equations leading to the
dispersion relation and the comparison with other simpler cases.
Conclusions are presented in Section III.

\section{Acoustic radial modes in cylinders and cylindrical shells of general
anisotropic crystals}

We shall consider a general anisotropic crystal. The matrix of elastic coefficients
will by given in this case by

\begin{eqnarray}
C_{\alpha \beta} & = & \left[\begin{array}{cccccc}
C_{11} & C_{12} & C_{13} & C_{14} & C_{15} & C_{16} \\
C_{12} & C_{22} & C_{23} & C_{24} & C_{25} & C_{26} \\
C_{13} & C_{23} & C_{33} & C_{34} & C_{35} & C_{36} \\
C_{14} & C_{24} & C_{34} & C_{44} & C_{45} & C_{46} \\
C_{15} & C_{25} & C_{35} & C_{45} & C_{55} & C_{56} \\
C_{16} & C_{26} & C_{36} & C_{46} & C_{56} & C_{66}
\end{array}\right] 
\end{eqnarray}

including twenty-one independent constants.

We shall consider a homogeneous thick cylinder of infinite length
and radius $R$, or a homogeneous cylindrical shell of infinite
length, inner radius $a$, outer radius $b$ and thickness $h$ ($b=a+h$).
We shall use the cylindrical coordinate system ($r$,$\theta$,$z$).

We assume that the mass density $\rho$ is constant in the cylinder
(cylindrical shell). The cylinder axis coincides with the
crystalline axis $z$ and the surfaces are stress free.

The strain-displacement relations are

\begin{eqnarray}
\epsilon_{rr}=\displaystyle{\frac{\partial u_{r}}{\partial r}} \;;
& \epsilon_{\theta\theta}=\displaystyle{\frac{1}{r}}
\left(\displaystyle{\frac{\partial u_{\theta}}{\partial
\theta}}+u_{r}\right)\;; & \epsilon_{zz}=\displaystyle{\frac{\partial u_{z}}{\partial z}}\;; \nonumber \\
\epsilon_{r\theta}=\displaystyle{\frac{1}{2}}\left(\displaystyle{\frac{1}{r}}
\displaystyle{\frac{\partial u_{r}}{\partial \theta}}+
\displaystyle{\frac{\partial u_{\theta}}{\partial r}}-
\displaystyle{\frac{u_{\theta}}{r}}\right)\;; & \epsilon_{\theta
z}=\displaystyle{\frac{1}{2}} \left(\displaystyle{\frac{\partial
u_{\theta}}{\partial z}}+ \displaystyle{\frac{\partial
u_{z}}{\partial \theta}}\right)\;; &
\epsilon_{rz}=\displaystyle{\frac{1}{2}}\left(\displaystyle{\frac{\partial
u_{r}}{\partial z}}+ \displaystyle{\frac{\partial u_{z}}{\partial
r}}\right)\;.
\end{eqnarray}

The equations of motion are given now by

\begin{eqnarray}
\displaystyle{\frac{\partial\sigma_{rr}}{\partial r}}+\displaystyle{\frac{1}{r}
\frac{\partial\sigma_{r\theta}}{\partial\theta}}+\displaystyle{\frac{\partial\sigma_{rz}}{\partial z}}+
\displaystyle{\frac{1}{r}}(\sigma_{rr}-\sigma_{\theta\theta}) & = & \rho
\displaystyle{\frac{\partial^{2}u_{r}}{\partial t^{2}}} \nonumber \\
\displaystyle{\frac{\partial\sigma_{r\theta}}{\partial r}}+\displaystyle{\frac{1}{r}
\frac{\partial\sigma_{\theta\theta}}{\partial\theta}}+\displaystyle{\frac{\partial\sigma_{\theta z}}{\partial z}}+
\displaystyle{\frac{2}{r}}\sigma_{r\theta} & = & \rho
\displaystyle{\frac{\partial^{2}u_{\theta}}{\partial t^{2}}} \nonumber \\
\displaystyle{\frac{\partial\sigma_{rz}}{\partial r}}+\displaystyle{\frac{1}{r}
\frac{\partial\sigma_{\theta z}}{\partial\theta}}+\displaystyle{\frac{\partial\sigma_{zz}}{\partial z}}+
\displaystyle{\frac{1}{r}}\sigma_{rz} & = & \rho
\displaystyle{\frac{\partial^{2}u_{z}}{\partial t^{2}}} \;,
\end{eqnarray}

where

\begin{equation}
\bm{\sigma}=\bm{C}\cdot\bm{\epsilon} \,.
\end{equation}

The general solution of this problem, as we told before, must be
performed numerically, but we shall consider now the particular
case of the radial modes.

The radial modes are purely radial vibrations, so that $u_{r}\neq
0$ and $u_{\theta}=u_{z}=0$. In the same way

\begin{equation}
\displaystyle{\frac{\partial u_{r}}{\partial\theta}}=
\displaystyle{\frac{\partial u_{r}}{\partial z}}=0 \;.
\end{equation}

It can be shown that the second and third equations in (3) are
satisfied automatically and we are left with

\begin{equation}
\displaystyle{\frac{d^{2}u_{r}}{dr^{2}}}+\displaystyle{\frac{1}{r}\frac{du_{r}}{dr}}+
(\beta^{2}_l-\displaystyle{\frac{\mu^{2}}{r^{2}}})u_{r}=0 \;,
\end{equation}

with $\beta^{2}_{l}=\displaystyle{\frac{\rho\omega^{2}}{C_{11}}}$
and $\mu^{2}= \displaystyle{\frac{C_{22}}{C_{11}}}$.

Eq.(6) is a Bessel equation of non-integer order. Thus we have the solution

\begin{eqnarray}
u_{r}(r) & = & \left\{ \begin{array}{lll}
AJ_{\mu}(\beta_{l}r)+BJ_{-\mu}(\beta_{l}r) & , & \beta_{l}>0 \\
Ar+ \displaystyle{\frac{B}{r}} & , & \beta_{l}=0 \;.
\end{array}
\right.
\end{eqnarray}

The boundary conditions at the surfaces are now

\begin{equation}
C_{11}\frac{du_{r}}{dr}+C_{12}\frac{u_{r}}{r}=0 \;.
\end{equation}

In the case of a thick cylinder, in order that $u_{r}$ be finite at
$r$=0, $B$=0. Thus we shall have:

\begin{eqnarray}
u_{r}(r) & = & \left\{ \begin{array}{lll}
AJ_{\mu}(\beta_{l}r) & , & \beta_{l}>0 \\
Ar & , & \beta_{l}=0 \;.
\end{array}
\right.
\end{eqnarray}

If $\sigma_{rr}(R)$=0, $A$=0 when $\beta_{l}$=0, and when
$\beta_{l}>$ 0 we have

\begin{equation}
C_{11}J^{\prime}_{\mu}(\beta_{l}R)+C_{12}\frac{1}{R}J_{\mu}(\beta_{l}R)=0 \;.
\end{equation}

As

\begin{equation}
J^{\prime}_{\mu}(\beta_{l}r)=\beta_{l}J_{\mu-1}(\beta_{l}r)-\frac{\mu}{r}J_{\mu}(\beta_{l}r)
\end{equation}

eq.(10) becomes

\begin{equation}
C_{11}\beta_{l}J_{\mu-1}(\beta_{l}R)+(C_{12}-C_{11}\mu)\frac{1}{R}J_{\mu}(\beta_{l}R)=0 \;,
\end{equation}

thus giving

\begin{equation}
\beta_{l}RJ_{\mu-1}(\beta_{l}R)=(\mu-\frac{C_{12}}{C_{11}})J_{\mu}(\beta_{l}R) \;.
\end{equation}

This equation is a generalization of the expression obtained for an infinitely
long isotropic cylinder \cite{29}

\begin{equation}
\beta_{l}J_{0}(\beta_{l}R)=\frac{1}{R}\frac{(1-2\nu)}{1-\nu}J_{1}(\beta_{l}R)
\end{equation}

with the Poisson's factor
$\nu=\displaystyle{\frac{C_{12}}{2(C_{12}+C_{44})}}$.

In the isotropic case $C_{11}=C_{22}$ and $C_{11}=C_{12}+2C_{44}$.
Thus it is easy to see that eq.(13) becomes in that case eq.(14).

In the case of the cylindrical shell, by using the boundary condition given in eq.(8), we obtain:

\begin{eqnarray}
A(C_{11}J^{\prime}_{\mu}(\beta_{l}r)+C_{12}\frac{1}{r}J_{\mu}(\beta_{l}r))+
B(C_{11}J^{\prime}_{-\mu}(\beta_{l}r)+C_{12}\frac{1}{r}J_{-\mu}(\beta_{l}r)) & , &
\beta_{l}>0 \nonumber  \\
A(C_{11}-C_{12})-B(C_{11}-C_{12})\frac{1}{r^{2}} & , & \beta_{l}=0 \;.
\end{eqnarray}

As $\sigma_{rr}(a)=\sigma_{rr}(b)$=0, we must have $A=B$=0, for $\beta_{l}$=0, whereas for $\beta_{l}>$0
we have

\begin{eqnarray}
\left| \begin{array}{cc}
C_{11}\beta_{l}J^{\prime}_{\mu}(\beta_{l}a)+C_{12}\displaystyle{\frac{1}{a}}J_{\mu}(\beta_{l}a)\;\;\;
&
C_{11}\beta_{l}J^{\prime}_{-\mu}(\beta_{l}a)+C_{12}\displaystyle{\frac{1}{a}}J_{-\mu}(\beta_{l}a)
\\
\\C_{11}\beta_{l}J^{\prime}_{\mu}(\beta_{l}b)+C_{12}\displaystyle{\frac{1}{b}}J_{\mu}(\beta_{l}b)\;\;\;
&
C_{11}\beta_{l}J^{\prime}_{-\mu}(\beta_{l}b)+C_{12}\displaystyle{\frac{1}{b}}J_{-\mu}(\beta_{l}b)
\end{array} \right| =0 \;.
\end{eqnarray}

This is the dispersion relation for the elastic radial modes of an
infinitely long cylindrical shell in the case of general
anisotropy. In hexagonal crystals $C_{22}=C_{11}$ and $C_{12}=
C_{11}-2C_{66}$. Then the last equation reduces to eq.(33) in
Ref.14, for the hexagonal crystals.

In the case of a thin shell it is possible to make an expansion to
first order in $\displaystyle{\frac{h}{a}}$ in eq.(16). Then we
arrive to

\begin{equation}
[(C^{2}_{12}-C_{11}C_{22})+C^{2}_{11}\beta^{2}_{l}a^{2}]\beta_{l}a[2\mu
J_{\mu}(\beta_{l}a)
J_{-\mu}(\beta_{l}a)+\beta_{l}aJ_{\mu}(\beta_{l}a)J_{-\mu-1}(\beta_{l}a)-\beta_{l}a
J_{\mu-1}(\beta_{l}a)J_{-\mu}(\beta_{l}a)]=0\;.
\end{equation}

Taking into account that

\begin{eqnarray}
J^{\prime}_{\mu}(\beta_{l}a) & = & \beta_{l}J_{\mu-1}(\beta_{l}a)-\frac{\mu}{a}J_{\mu}(\beta_{l}a) \nonumber \\
J^{\prime}_{-\mu}(\beta_{l}a) & = &
\beta_{l}J_{-\mu-1}(\beta_{l}a)+\frac{\mu}{a}J_{-\mu}(\beta_{l}a)
\end{eqnarray}

it can be seen that the factor involving the Bessel functions in
eq.(17) is the Wronskian, having the value

\begin{equation}
J_{\mu}(\beta_{l}a)J^{\prime}_{-\mu}(\beta_{l}a)-J^{\prime}_{\mu}(\beta_{l}a)J_{-\mu}(\beta_{l}a)=
-\displaystyle{\frac{2\sin(2\pi)}{\pi\beta_{l}a}}\neq 0 \;.
\end{equation}

Thus the dispersion relation for the lowest breathing mode of a
thin cylindrical shell of a general anisotropic crystal is given
by

\begin{equation}
(C^{2}_{12}-C_{11}C_{22})+C^{2}_{11}\beta^{2}_{l}a^{2}=0 \;,
\end{equation}

which can be put in the form

\begin{equation}
\omega=\displaystyle{\frac{1}{a}\sqrt{\frac{C_{11}C_{22}-C^{2}_{12}}{\rho
C_{11}}}= \frac{2}{d}\sqrt{\frac{C_{11}C_{22}-C^{2}_{12}}{\rho
C_{11}}}} \;,
\end{equation}

$d=2a$ being the diameter.

In the case of hexagonal crystals $C_{22}=C_{11}$ and $C_{12}=
C_{11}-2C_{66}$ and eq.(21) reduces to

\begin{equation}
\omega=\frac{2}{a}\sqrt{\frac{C_{66}(C_{12}+C_{66})}{\rho(C_{12}+2C_{66})}}
\end{equation}

given in Ref.14.

For an isotropic crystal $C_{22}=C_{11}$ and $C_{12}=
C_{11}-2C_{44}$. Then eq.(21) reduces to

\begin{equation}
\omega=\frac{2}{a}\sqrt{\frac{C_{44}(C_{12}+C_{44})}{\rho(C_{12}+2C_{44})}}
\end{equation}

which coincides with the expressions given in Refs.11,14.

Eq.(21) is a simple analytic expression valid for all crystal systems. It
provides in a quick way a first estimate of the lowest breathing
mode frequency for nanowires. It was shown in Ref.14 that
eq.(22) gave extremely good agreement with experimental and first
principles theoretical values of the lowest breathing mode
frequency of different nanotubes. It can be expected that the same 
will happen for eq.(21) when applied to other materials and 
crystal systems.

Eq.(16) gives the frequencies of higher order breathing modes.

\section{Conclusions}

We have obtained expressions in closed form for the dispersion relation of radial modes in
infinitely long cylinders and cylindrical shells of general anisotropic materials. When
considering the case of a very thin shell we arrive to a very simple analytic expression
for the lowest breathing mode. This expression covers the cases of isotropic and hexagonal
systems, previously given in the literature.

\acknowledgments This work was partially supported by the Spanish
Ministerio de Econom\'{\i}a y Competitividad under Grant MAT2009-14578-C03-03. 
D. M.-G. acknowledges financial support from the FPI Program of the 
Spanish Ministerio de Econom\'{\i}a y Competitividad.

\newpage

\end{document}